\documentclass[12pt]{article}
\usepackage{graphicx}

\usepackage{geometry}
\usepackage{setspace}
\usepackage{graphicx}
\usepackage{amssymb}
\usepackage{epstopdf}
\usepackage{multirow}
\usepackage{extarrows}
\usepackage{rotating}
\usepackage{dsfont}
\usepackage{bm}
\usepackage{empheq}
\usepackage{endnotes}
\newcommand\mathC{\mkern1mu\raise2.2pt\hbox{$\scriptscriptstyle|$}
        {\mkern-7mu\rm C}}              

\title{Discerning ``indistinguishable" quantum systems} 
          \author{Adam Caulton\footnote{c/o The Faculty of Philosophy, University of Cambridge, Sidgwick Avenue, Cambridge, UK, CB3 9DA. Email: aepw2@cam.ac.uk}}
             \date{}
             
\begin{document}
\maketitle

\begin{abstract}

In a series of recent papers, Simon Saunders, Fred Muller and Michael Seevinck have collectively argued, against the folklore, that some non-trivial version of Leibniz's principle of the identity of indiscernibles is upheld in quantum mechanics.  They argue that \emph{all} particles---fermions, paraparticles, anyons, even bosons---may be weakly discerned by some physical relation.  Here I show that their arguments make illegitimate appeal to non-symmetric, i.e.~permutation-non-invariant, quantities, and that therefore their conclusions do not go through.  However, I show that alternative, symmetric quantities may be found to do the required work.  I conclude that the Saunders-Muller-Seevinck heterodoxy can be saved after all.

\end{abstract}


\tableofcontents

\section{Introduction}

\subsection{Getting clear on Leibniz's Principle}

What is the fate of Leibniz's Principle of the Identity of Indiscernibles for quantum mechanics?  It depends, of course, on how the Principle is translated into modern (enough) parlance for the evaluation to be made.  Modern logic provides a framework in which some natural regimentations may be articulated, which, even if they would not have been of interest to Leibniz's original project, are nevertheless worthy of investigation in their own right.

One informal gloss of the Principle is that no two objects share all the same properties.  Grant that we may regiment by taking `object' in the Fregean-Quinean sense of an occupant of the first-order domain.  Then what might count as a property?  If, for each first-order model, we universally quantify over the interpretations afforded in that model to the distinguished predicates, then the Principle is contingent.  That is, in some models the non-logical vocabulary is expressive enough to define identity; in others it is not. But this regimentation might be thought to make a metaphysical principle too much a hostage to the fortunes of language. Why only quantify over the properties for which we have the corresponding predicates?

An alternative is to regiment the Principle so that, in each model, one generalises over  all subsets of the first-order domain. The result is a (second-order) logical truth, since for any object $a$ there is its singleton $\{a\}$.  But here I see at least two objections.  First, sets are not properties.  But no worries: for any subset of the domain, there is at least one property to which it corresponds---namely, the property of belonging to that subset.  Thus generalisation over subsets may be taken as covert generalisation over \emph{these} properties at least, and discernibility by these properties entails discernibility simpliciter.  (All this, of course, only so long as there \emph{are} sets.)  The second, more serious, objection is that the singleton sets one discerns by are precisely as discernible as the objects that are their unique members.  In what sense, then, is it an achievement to have discerned those objects with those sets?  In other words: when the properties one quantifies over correspond to the subsets of the first-order domain, the Principle becomes trivial. That should come as no surprise---as I said, it is a logical truth---but it cannot be the regimentation of Leibniz's Principle that we are looking for.

The solution---or, at least, the solution I favour for the purposes of this paper---is to retreat to generalising over the interpretations assigned to the distinguished predicates, but suitably to relativise the Principle to the appropriate collection of properties and relations: namely, those for which the distinguished predicates stand.  The Principle now reads: ``No two objects share all the same properties expressible in the language.''  One may then, if it is so desired, recover an absolute version of the Principle by ensuring that each property or relation taken to exist has its corresponding predicate in the non-logical vocabulary---assuming such a thing possible. 

With this regimentation, the Principle may be given an explicit first-order logical form.  For each predicate, one forms a biconditional expressing co-satisfaction for $x$ and $y$.  If the predicate is 2-place or more, one universally generalises over the other argument places, and makes sure that there is a bi-conditional for each argument position of the predicate.   The conjunction of all such bi-conditionals is then asserted to be co-extensional with `$x=y$', and we have a (putative) explicit definition of identity.  The result is the Hilbert-Bernays (1934) axiom, made famous by Quine (e.g.~1960) and revived by Saunders (2003a, 2003b, 2006):
$$
\begin{array}{cc}
	\begin{array}{rcll}
		\forall x\forall y \bigg\{x = y &\equiv& \Big[\ \ldots \ \wedge \
		(F_ix \equiv F_iy)
		\ \wedge\ \ldots\\\\
		&&\quad \ldots \ \wedge \ \forall z \left(
		(G_jxz \equiv G_jyz) \wedge
		(G_jzx \equiv G_jzy)\right)
		\ \wedge \ \ldots\\\\
		&&\quad \ldots \ \wedge \ \forall z \forall w \Big(
		(H_kxzw \equiv H_kyzw) \wedge
		(H_kzxw \equiv H_kzyw)
		\\ &&\qquad\qquad\qquad\qquad\qquad\qquad \wedge \
		(H_kzwx \equiv H_kzwy)\Big)
		\ \wedge \ \ldots \Big]\bigg\}
	\end{array}
	&
(HB)
\end{array}
$$
Of course, one must assume that there are finitely many properties and relations, unless one cares to appeal to infinitary languages or some form of parameterization  (cf.~Caulton and Butterfield 2011, \S2.1).  The question whether the Principle is true in any theory can now be made precise.  If one takes a theory to be a set $T$ of sentences, the question is: Does $T$ logically entail ($HB$)?  If one takes a theory to be a set $\mathfrak{M}$ of models, the question is:  Is ($HB$) true in every model in $\mathfrak{M}$?

\subsection{The folklore}\label{PIIold}

Let $T =$ quantum mechanics, or $\mathfrak{M} =$ the models of quantum mechanics.  Is ($HB$) a logical consequence of $T$, or true in all models in $\mathfrak{M}$---at least when the first-order variables are restricted to quantum particles?
Until about eight years ago, the folklore has been
that quantum particles  cannot be discerned, so that Leibniz's Principle \emph{fails}. 

To explain this in more detail, it will be clearest to start with an even earlier folklore, inherited from the founders of quantum mechanics. This folklore has it that Pauli's exclusion principle for fermions---or better: symmetrization for bosons and anti-symmetrization for fermions---means that:
\begin{enumerate}
\item[(a)] bosons can be in the same state; but
 \item  [(b)] fermions cannot be;
so that 
\item [(c)] Leibniz's Principle holds for fermions but not bosons.
\end{enumerate}
(For an expression of these three views, see e.g.~Weyl 1928, 241.) In fact, these claims can and should be questioned. Under scrutiny, and certain interpretative assumptions, each of (a) to (c) fail, and it seems that Leibniz's Principle is pandemically {\em false} in quantum mechanics.

For first: under the standard interpretation of the formalism,\footnote{I will not question this widespread interpretation of the formalism here, although, like Earman (ms.) and Dieks and Lubberdink (2011), I am greatly suspicious of it.} any two bosons or any two fermions of the same species  are \emph{absolutely indiscernible}, in the sense that no quantity (``observable'') exists which can discern them.\footnote{For a discussion of absolute discernibility, see e.g.~Saunders (2003b), Muller and Saunders (2008), and Caulton and Butterfield (2011).}   For any assembly of  fermions or bosons, and any state of that assembly (appropriately
(anti-) symmetrized), and any two particles in the assembly, the two particles' probabilities for all single-particle quantities are equal; and so are appropriate corresponding two-particle conditional probabilities, including probabilities using conditions about a third constituent.  In more technical language: 
according to the usual procedure of extracting the reduced density operator of a particle by tracing out the states for all the other particles in the assembly, we obtain the result that for all (anti-) symmetrized states of the assembly, one obtains \emph{equal} reduced density operators for every particle. (Margenau 1944; French \& Redhead 1988; Butterfield 1993; Huggett 1999; 2003; Massimi 2001; French \& Krause 2006, 150-73.)

Thus not only can fermions `be in the same state', just as much as bosons can be---{\em pace} the informal slogan form of Pauli's exclusion principle---also, a pair of indistinguishable particles of either species {\em must} be in the same state.  This result \emph{appears} to entail that Leibniz's Principle is pandemically {\em false} in quantum theory.\footnote{It may be of interest to note that these results can also be shown to hold for paraparticles, so long as one follows Messiah and Greenberg's (1964) recommendation of working with `generalised rays' (i.e.~multi-dimensional subspaces) instead of one-dimensional rays; see Huggett (2003).}

\subsection{A new folklore?}

Such was the folklore until eight years ago. But this folklore has {also} recently been called into question by Simon Saunders, Fred Muller and Michael Seevinck.  For (building on the Hilbert-Bernays account of identity) there are ways of distinguishing particles that outstrip the notion of a quantum state for a particle (and its assocated probabilities, including conditional probabilities)---and yet which {\em are} supported by quantum theory.  That is: the folklore has overlooked predicates on the right-hand side of ($HB$) which may, after all, sanction the right-to-left implication.

For as the Hilbert-Bernays account teaches us, two objects can be discerned even if they share all their monadic properties and their relations to all other objects---and even if any relation that they hold to one another is held symmetrically. That is: they can be discerned \emph{weakly}.\footnote{The terminology originates with Quine (1960).}  Thus if, for some relation $R$ and two objects $a$ and $b$, we have that $Rab$ and $Rba$, then $a$ and $b$ must be distinct if either $Raa$ or $Rbb$ (or both) fails.

It remains to provide such a relation that is legitimate within quantum mechanics.  This task was undertaken in its most general form for fermions by Muller and Saunders (2008), and for all particles by Muller and Seevinck (2009). (This work built upon an original suggestion by Saunders (2003b), which took inspiration from the fact that two particles in the spin singlet state may be said to have opposite spin (or to have vanishing combined total spin) without having to pick a spin direction.)  In the following two sections (Sections \ref{MullSau} and \ref{MullSee}), I will  appraise the results in these two papers.  I will conclude that Saunders, Muller and Seevinck were largely correct in their general conclusion that weakly discerning relations may be found, but that their proofs make incorrect assumptions---incorrect, that is, on their own terms---about which aspects of the quantum formalism represent genuine physical structure.  I will propose a friendly amendment to the Saunders-Muller-Seevinck results in Section \ref{SMSamend}, and secure the fact that particles are always \emph{weakly} discernible, whether they be bosons, fermions or paraparticles.

\section{Muller and Saunders on discernment} \label{MullSau}

\subsection{The Muller-Saunders result}

Here I briefly present the main result contained in Muller and Saunders (2008).  First I follow these authors in establishing three important distinctions in the way that particles may be discerned.
\begin{enumerate}
\item \emph{Absolute vs.~relative vs.~weak discernment.}  The first distinction relates to the logical form of the predicates used to discern the particles.  As we have seen, all fermions and bosons are absolutely indiscernible; they are also relatively indiscernible.  Thus our only hope is to discern them \emph{weakly}.  
\item \emph{Mathematical vs.~physical discernment.}  Of course, it is crucial that the properties and relations used to discern the particles be \emph{physical}: we cannot appeal to elements of the theory's mathematical formalism which have no representational function.  Thus, for example, we cannot discern two particles in an assembly merely by appealing to the fact that the Hilbert space for that assembly is a tensor product of two copies of a factor Hilbert space.  For all we know, this representative structure may be redundant; there may in fact only be one particle.  So we must instead appeal to quantities in the formalism which genuinely represent \emph{physical} quantities.  Like Muller, Saunders and Seevinck, I call this sort of legitimate discernment `physical discernment'.  I call instances of spurious discernment `mathematical discernment'---Muller and Saunders instead use the phrase `lexicon discernment'; but it is important to distinguish between mathematical objects (like Hilbert spaces) and mathematical language.  Thus I restrict ($HB$) above to contain only physical predicates; mathematical predicates (such as set membership: `$\in$') are not to be included.
\item \emph{Categorical vs.~probabilistic discernment.}  The final distinction relates to the assumptions required to secure the discernment.  Muller and Saunders call an instance of discernment `categorical' just in case it requires no appeal to the Born rule, and `probabilistic' otherwise.  The main advantage of categorical, as opposed to probabilistic, discernment is that by by-passing probabilistic notions its validity need not wait on any solution to the quantum measurement problem.  However, this advantage is in my view only slight, since surely any solution to the measurement problem must anyway secure at least an approximate vindication of the Born rule.   Here the restriction is not on ($HB$) but the theory taken to entail it.  Categorical discernment means entailment by quantum mechanics \emph{without} the Born rule as a postulate. 
\end{enumerate}

We are now in a position to state the main Muller-Saunders result:

\begin{tabular}{cl}
(SMS1) & Fermions are categorically, weakly, physically discernible.
\end{tabular}

\emph{Reconstruction of proof} (cf.~Muller and Saunders 2008, 536): We consider an assembly of only two fermions, so our Hilbert space is $\mathcal{A}(\mathcal{H}\otimes\mathcal{H})$; the result is easily extendible for more than two particles (cf.~Muller and Saunders 2008, 534).  Select some complete set of projection operators $\{E_i\}, \sum_i E_i = \mathds{1}$, for the single-particle Hilbert space $\mathcal{H}$ and define $P_{ij} := E_i - E_j$. Then define $P_{ij}^{(1)} := P_{ij}\otimes\mathds{1}$ and  $P_{ij}^{(2)} := \mathds{1}\otimes P_{ij}$, where the superscripts are \emph{labels} for particles 1 and 2.  We then define the following relation:
\begin{equation} \label{defR1}
R_t(x, y) \quad \mbox{iff} \quad \sum_{i,j} P^{(x)}_{ij}P^{(y)}_{ij}\rho= t\rho ,
\end{equation}
where $t\in \mathbb{R}$, $\rho$ is the density operator representing the state of the assembly, and the indices $i,j$ range over the projectors $E_i$.  

First we prove that 1 and 2 are categorically and weakly discerned by $R_t$ for some value of $t$.  To see that the discernment is categorical, it can be shown (cf.~Muller and Saunders 2008, 533) that, with dim$(\mathcal{H})\geqslant 2$, for every state $|\Psi\rangle\in\mathcal{A}(\mathcal{H}\otimes\mathcal{H})$,
\begin{equation}
\sum_{i,j}P^{(1)}_{ij}P^{(2)}_{ij}|\Psi\rangle = \sum_{i,j}P^{(2)}_{ij}P^{(1)}_{ij}|\Psi\rangle = -2|\Psi\rangle
\end{equation}
and
\begin{equation}
\sum_{i,j}\left(P^{(1)}_{ij}\right)^2|\Psi\rangle = \sum_{i,j}\left(P^{(2)}_{ij}\right)^2|\Psi\rangle = 2(d-1)|\Psi\rangle\ ,
\end{equation}
where $d = \mbox{dim}(\mathcal{H}) $.  Thus every state of the assembly is an eigenstate of the operators used in the definition of $R_t$; and so we do not need to assume the Born rule.  $R_t$  therefore promises to provide  categorical discernment.

To see that $R_t$ discerns the particles weakly for some $t$, note that $R_t(1,1)$ and $R_t(2,2)$ iff $t=2(d-1)$, whereas $R_t(1,2)$ and $R_t(2,1)$ only if $t=-2$.\footnote{Remember that `1' and `2' serve as particle labels in the expressions `$R_t(1,2)$', etc.}  So the relations $R_{2(d-1)}$ and $R_{-2}$ both serve to weakly discern particles 1 and 2.

Finally, it remains to be shown that $R_t$ is a \emph{physical} relation.  I turn to Muller and Saunder's criteria (2008, pp.~527-8):
\begin{quote}
(Req1) \emph{Physical meaning.} All properties and relations should be transparently defined in terms of physical states and operators that correspond to physical magnitudes, as in [the weak projection postulate],\footnote{The weak projection postulate is effectively Einstein, Podolsky and Rosen's (1935) reality condition that the assembly's being in an eigenstate of any self-adjoint operator $Q$ with eigenvalue $q$ is a sufficient condition for the assembly's possessing the property corresponding to the quantity's $Q$ having value $q$.  This is an interpretative principle, which, like Muller and Saunders (2008) and Muller and Seevinck (2009), I take for granted.\label{EPR1}} in order for the properties and relations to be physically meaningful. 

(Req2) \emph{Permutation invariance.} Any property of one particle is a property of any other; relations should be permutation-invariant, so binary relations are symmetric and either reflexive or irreflexive. 
\end{quote}
(Req2) is clearly true of $R_t$.  (Req1) is also true of $R_t$, provided that: (i) the projectors $E_i$ are physically meaningful; and (ii) the physical meaningfulness of operators is preserved under mathematical operations; for our purposes these must include: arithmetical operations, i.e. addition and multiplication; and tensor multiplication with the identity.  (Note: Muller and Saunders take (i) (along with (Req2)) to be sufficient to establish that $R_t$ is a physical relation (2008, 534-5).  However, it is clear that (ii) is also required.) $\Box$

\subsection{Commentary on the Muller-Saunders proof}

I take no issue with Muller and Saunders' claim that their relations $R_t$ provide categorical and weak discernment.  However, I question whether the relations $R_t$ may properly be considered physical.  I take no issue with the idea that projectors \emph{per se} are physically meaningful (like Muller and Saunders, I agree that these can be considered to represent specific experimental questions with a yes/no answer); but $R_t$ is defined in terms of \emph{non-symmetric} projectors $E_i \otimes \mathds{1}$, etc.  And it is compulsory---i.e.,~a necessary condition for representing a physical quantity---that the quantities obey the Indistinguishability Postulate (IP), which demands that all physical quantities be permutation invariant. (Cf.~Messiah and Greenberg 1964.)

This brings us to my criticism of (Req2), which has two components.  First: it misapplies the correct idea that physical quantities must be symmetric.  By requiring only that the {relations} defined \emph{from} the quantum mechanical quantities be symmetric, (Req2) fails to rule out use of quantum mechanical quantities which may themselves be non-symmetric.  To take a simple illustration of this point: `$x$ is particle 1 and $y$ is particle 2' clearly fails to be a physical relation, both in the proper sense, and in terms of (Req2).  But the relation `$x$ is particle 1 and $y$ is particle 2, or $x$ is particle 2 and $y$ is particle 1' is equally unphysical, yet it does satisfy (Req2).

It may be replied that this is where (Req1) comes in.  But this brings us to the second component of my criticism of (Req2): it is redundant.  For it is anyway necessary for a quantity to be symmetric to satisfy (Req1), since any non-symmetric quantity contravenes IP, and therefore cannot represent a `physical magnitude'.  Indeed: since (Req1) already demands that the quantities be physical, why do we need \emph{any} further requirement?

It may be objected on behalf of Muller and Saunders that, while the quantities $P^{(1)}_{ij}$ and $P^{(2)}_{ij}$ indeed fail to be symmetric, the quantities defined in terms of them---namely, the  $\sum_{i,j} P^{(x)}_{ij}P^{(y)}_{ij}$---\emph{are} symmetric.  This is indeed true: $\sum_{i,j} \left(P^{(1)}_{ij}\right)^2 = \sum_{i,j} \left(P^{(2)}_{ij}\right)^2 = 2(d-1)\mathds{1}\otimes\mathds{1}$ and $\sum_{i,j} P^{(1)}_{ij}P^{(2)}_{ij} = \sum_{i,j} P^{(2)}_{ij}P^{(1)}_{ij} = 2(\sum_iE_i\otimes E_i - \mathds{1}\otimes\mathds{1})$, where $\mathds{1}$ is the identity on $\mathcal{H}$.  (Note that the restrictions of both quantities to the anti-symmetric sector, $\mathcal{A}(\mathcal{H}\otimes\mathcal{H})$, are  multiples of the identity on that sector.)   But I see no force in the objection: the physical significance of these quantities was supposed to rest on their being constructions out of quantities like $E_i \otimes\mathds{1}$; yet it is precisely these quantities which run afoul of IP.  

Without any convincing account of the physical significance of the building blocks of the  $\sum_{i,j} P^{(x)}_{ij}P^{(y)}_{ij}$, these quantities must be assessed for their physical significance on their own terms.  But, since they are all multiples of the identity on the assembly's state space, this significance is trivial: they all correspond to experimental questions which yield the same answer on every physical state.\footnote{I am very grateful to Nick Huggett for discussions about this point.}

This triviality is a problem for Muller and Saunders, since it blocks the $R_t$ from being  physical relations.  If we now attempt to redefine the $R_t$ in a way that avoids misleading reference to the fraudulently physical quantities $P^{(x)}_{ij}$ we obtain:
\begin{equation}
R_t(x, y) \quad \mbox{iff} \quad  (x=y \mbox{ and } 2(d-1)\rho= t\rho) \mbox{ or } (x\neq y \mbox{ and } (-2)\rho= t\rho) 
\end{equation}
This is equivalent to:
\begin{equation}\label{defR2}
R_t(x, y) \quad \mbox{iff} \quad  (x=y \mbox{ and } t = 2(d-1)) \mbox{ or } (x\neq y \mbox{ and } t=-2) .
\end{equation}
So long as we have a definition of the $R_t$ in terms of quantities that seems (i.e.~from the point of view of the syntax) to treat the $x=y$ and $x\neq y$ cases equally, the fact that a \emph{different} quantity (i.e.~a different multiple of the identity) underlies each of these two cases is tolerable.  (In just the same way, $Rxy$ and $Rxx$ are strictly speaking different predicates---since one refers to a relation while the other refers to a monadic property---yet it is normal to treat any instance of  $Rxx$ as a special {instance} of $Rxy$.  Indeed: weak discernment \emph{relies} on this being  legitimate.)  But since the quantities $\sum_{i,j} P^{(x)}_{ij}P^{(y)}_{ij} $ must be taken at face value---that is, as nothing but multiples of the identity---we must adopt definition (\ref{defR2}) over definition (\ref{defR1}), and definition (\ref{defR2}) is hopelessly gerrymandered and unphysical.  Thus Muller and Saunders' proof that any two fermions are \emph{physically} discernible does not go through.

In Section \ref{SMSamend}, I propose an alternative relation which will discern fermions physically and weakly, though not categorically.  But first let me address the main results in Muller and Seevinck (2009).

\section{Muller and Seevinck on discernment} \label{MullSee}

\subsection{The Muller-Seevinck result}

Muller and Seevinck use a similar framework to Muller and Saunders (2008): specifically, they carry over the three distinctions between kinds of discernment presented above, and the two requirements for physical significance, (Req1) and (Req2).\footnote{Muller and Seevinck (2009, pp.~185-6) entertain adding a third requirement, to the effect that discernment by a relation is `authentic' only if it is irreducible to monadic properties, and discernment by a monadic property is `authentic' only if it is irreducible to relations.  They reject this extra requirement, as do I; but my reason is different.  My reason is that physical meaning (embodied in (Req1)) is all one could, and should, reasonably ask for---so long as that is taken to entail the requirement that the Indistinguishability Postulate is satisfied; cf.~my commentary of Muller and Saunders' proof in Section \ref{MullSau}.}  There are two main results to discuss: the first concerns spinless particles with infinite-dimensional Hilbert spaces; the second concerns spinning systems with finite-dimensional Hilbert spaces.
 
 I begin with their Theorem 1.  (Note that I rephrase their Theorems; cf.~Muller and Seevinck 2009, 189.)

\begin{tabular}{cp{12cm}}
(SMS2) & In an assembly with Hilbert space $\bigotimes^N L^2(\mathbb{R}^3)$ and the associated algebra of quantities $\mathcal{B}(\bigotimes^N L^2(\mathbb{R}^3))$, any two particles are categorically, weakly, physically discernible.
\end{tabular}

\emph{Reconstruction of proof} (cf.~Muller and Seevinck 2009, 189):  Again, for simplicity's sake, I restrict attention to the case of two particles ($N=2$).  Let $Q$ be the position operator for a single particle in some dimension (say $x$), and let be $P$ be the momentum operator in that same dimension.  (So $Q$ and $P$ are (partially) defined on $L^2(\mathbb{R}^3)$; and I shall not go into detail about the partialness of the domains of definition, which are adequately discussed by Muller and Seevinck.)  Now define $Q^{(1)} := Q\otimes\mathds{1}$ and $Q^{(2)} := \mathds{1}\otimes Q$, and   $P^{(1)} := P\otimes\mathds{1}$ and $P^{(2)} := \mathds{1}\otimes P$, where $\mathds{1}$ is the identity on $L^2(\mathbb{R}^3)$.

We may now define a relation $C$ as follows:
\begin{equation}
C(x,y) \quad \mbox{iff} \quad [P^{(x)}, Q^{(y)}]\rho = c\rho, \mbox{ for some } c \neq 0\ ,
\end{equation}
where $\rho$ is the density operator representing the state of the assembly.  Now for every state we have $C(1,1)$ and $C(2,2)$, since $[P^{(1)}, Q^{(1)}] = [P^{(2)}, Q^{(2)}] = -i\hbar\mathds{1}\otimes\mathds{1}$.  And we also have $\neg C(1,2)$ and $\neg C(2,1)$, since $[P^{(1)}, Q^{(2)}] =  [P^{(2)}, Q^{(1)}] = 0$.  Thus $C$  weakly discerns particles 1 and 2.  This discernment is categorical, since $C$ holds or not categorically, i.e.~without probabilistic assumptions.  And the discernment is physical, since $C$ meets (Req1) and (Req2). $\Box$

\emph{Commentary.}
First of all I note that the restriction in (SMS2), that each particles' state space be $L^2(\mathbb{R}^3)$ should count as no real restriction, since all real particles have spatial degrees of freedom.  Second: since the discernment is categorical, it is no restriction that the full (i.e.~un-symmetrized) Hilbert space is used in the proof: the proof carries over for all restrictions to symmetry sectors.

As in Section \ref{MullSau}, again I take no issue with the claim that the discernment is weak and categorical, but I do deny that it is physical.  The reason is the same as for Muller and Saunders (2008): namely, the proof uses unphysical quantities.  (Thus I deny that (Req1) is satisfied.)  Again we demand not just that the discerning relation be symmetric, but also that it be defined using only physical---\emph{a fortiori}, only symmetric---quantities.  And $Q^{(x)}$ and $P^{(x)}$, despite their tantalising intuitive interpretation, do not count as physical quantities.

I now turn to Muller and Seevinck's second main Theorem:

\begin{tabular}{cp{12cm}}
(SMS3) & In an assembly with a finite-dimensional Hilbert space $\bigotimes^N \mathbb{C}^{2s+1}$, where $s\in \{\frac{1}{2}, 1, \frac{3}{2}, \ldots\}$ and the associated algebra of quantities $\mathcal{B}(\bigotimes^N \mathbb{C}^{2s+1})$, any two particles are categorically, weakly, physically discernible using only their spin degrees of freedom.
\end{tabular}

\emph{Reconstruction of proof} (cf.~Muller and Seevinck 2009, 193-7):  Again I restrict attention to the case of two particles ($N=2$).  Let $\mathbf{S} = \sigma_x\mathbf{i} + \sigma_y\mathbf{j} + \sigma_k\mathbf{k}$ be the quantity representing a single particle's spin (so $\mathbf{S}$ acts on $\mathbb{C}^{2s+1}$).  Then we define $\mathbf{S}_1 := \mathbf{S}\otimes\mathds{1}$ and $\mathbf{S}_2 := \mathds{1}\otimes \mathbf{S}$, and the relation $T$ as follows:
\begin{equation} \label{Tdef}
T(x,y) \quad \mbox{iff} \quad \mbox{for all } \rho \in  \mathcal{D}(\mathbb{C}^{2s+1}\otimes \mathbb{C}^{2s+1}),\ |(\mathbf{S}_x + \mathbf{S}_y)|^2\rho = 4s(s+1)\hbar^2\rho.
\end{equation}
Recall that $|\mathbf{S}|^2 = s(s+1)\hbar^2\mathds{1}$; this entails that $|2\mathbf{S}_1 |^2 =  |2\mathbf{S}_2 |^2 = 4s(s+1)\hbar^2\mathds{1}\otimes\mathds{1}$; so $T(1,1)$ and $T(2,2)$ both hold.  Meanwhile, $|(\mathbf{S}_1 + \mathbf{S}_2)|^2 = |\mathbf{S}|^2\otimes\mathds{1} + \mathds{1}\otimes|\mathbf{S}|^2 + 2\mathbf{S}\otimes\mathbf{S} = 2s(s+1)\hbar^2\mathds{1}\otimes\mathds{1} +  2\mathbf{S}\otimes\mathbf{S}$.  But the eigenvalue spectrum of $|(\mathbf{S}_1 + \mathbf{S}_2)|^2 $ never exceeds $(2s)(2s+1)\hbar^2 < 4s(s+1)\hbar^2$, so $\neg T(1,2)$ and $\neg T(2,1)$ both hold.  This discernment is clearly weak.  It is categorical, since it relies on no probabilistic assumptions,  and it is physical, since $T$ satisfies (Req1) and (Req2).  $\Box$

\subsection{Commentary on the Muller-Seevinck result} 

I note that, in order to put the physical significance of $T$ on firmer ground, Muller and Seevinck extend the EPR reality condition (cf.~footnote \ref{EPR1}) to a necessary and sufficient condition, which they call the `strong property postulate'.   According to this postulate, the assembly possesses the property corresponding to the quantity's $Q$ having value $q$ \emph{if and only if} the assembly's state is an eigenstate of the self-adjoint operator $Q$, with eigenvalue $q$. This strengthening is required to establish that the assembly does not possess combined total spin $\sqrt{4s(s+1)}\hbar$ when it not in an eigenstate of the total spin operator.

Freedom from this stronger reality condition can be bought at the price of a concession to settle for probabilistic rather than categorical discernment.  For we may define the new relation $T'$:
\begin{equation} \label{T'def}
T'(x,y) \quad \mbox{iff} \quad  Tr\left(\rho|(\mathbf{S}_x + \mathbf{S}_y)|^2\right) = 4s(s+1)\hbar^2.
\end{equation}
It is clear that $T'$ discerns iff the ``de-modalized" version of $T$ discerns.  But the definition of $T'$ involves a commitment to the Born rule, so $T'$'s discernment is probabilistic.  This trade-off between the strong reality condition and the Born rule will also be a feature of my proposals in the following section.

  The previous objection I levelled against (SMS1) and (SMS2) \emph{appears} to be valid here too.  For, even though $ |(\mathbf{S}_1 + \mathbf{S}_2)|^2$ and $ |2\mathbf{S}_1 |^2 =|2\mathbf{S}_2 |^2$ are symmetric, once again their building blocks ($\mathbf{S}_1$ and $\mathbf{S}_2$) are not, and (it may be argued) it is only when defined in terms of these components that $T$ is not a gerrymandered relation.

However, my usual objection does \emph{not} hold in this case.  On the contrary, it seems reasonable  to take $T(x,y)$ as a natural physical relation, even though its explicit mathematical form depends on whether $x=y$ or $x\neq y$.  To see this, it should be enough that $T$ can be parsed in English as the relation: `the combined total spin of $x$ and $y$ has the magnitude $\sqrt{4s(s+1)}\hbar$ in all states'.  Combined total spin is a symmetric quantity, and it has obvious physical significance.  Therefore I do not take issue with the discerning relation being physical.

But I have two different objections in this case: one mild, the other more serious.  The mild objection is that the relation $T$ is different in a significant way from the previous relations $R_t$ and $C$.  While $R_t$ and $C$ both apply to a given state of the assembly, the definition of $T$ involves quantification over all states of the assembly.  It is therefore a \emph{modal} relation.  But appeal to modal relations in this context is problematic, since it threatens to trivialise the search for a discerning relation for every state.  It would turn out that Leibniz's Principle is necessarily true if it is possibly true: a result that is at best controversial (though Saunders 2003b seems to endorse it, taking ($HB$) as an explicit \emph{definition} of identity, as Quine 1960 also suggests).

(Note, incidentally, that the use of modal relations cannot be criticised on the grounds that it assumes haecceitism. It is natural---at least in standard practice---to use Hilbert space labels to cross-identify systems between states, and this seems to have a whiff of haecceitism about it.  However, this cross-identification strategy does  not entail haecceitism, since the quantification over states may be restricted to the (anti-) symmetric sectors, in which all states are already permutation-invariant, so that the issue of haecceitism is moot.)

This mild objection is easily met.  We simply drop the quantification over states in the definition of $T$.  If we do this, then the (unquantified) right-hand side of the definition (\ref{Tdef}) is still satisfied iff $x=y$, for all states $\rho$.  We thereby drop the modal involvement. Thus we define a new relation, to be parsed as `the combined total spin of $x$ and $y$ has the magnitude $\sqrt{4s(s+1)}\hbar$'.  The discernment remains categorical, since no probabilistic assumptions have been made.

The serious problem is that (SMS3) is only applicable to assemblies whose constituent particles have non-zero spin.  This might seem to be only a mild omission, since the only elementary spin-zero particle that actually exists  is the Higgs boson, and for a treatment of that we turn to quantum field theory.  However, it would be nice to establish the discernibility of quantum particles for \emph{all} values of spin, not just for the sake of the Higgs boson, but for the sake of any hypothetical particle,  actual or merely possible.

To sum up: the same problem beleaguers the first two results (SMS1) and (SMS2), which aim to demonstrate the discernibility of (respectively) fermions and any particles with spatial degrees of freedom.  The problem is that they both appeal to quantities which, in virtue of contravening IP, are non-physical.  The third result, (SMS3), avoids this problem (modulo dropping some unnecessary modal involvement).  However, it does not apply to particles with zero-spin.  I now turn to my proposal for discerning any species of particle, for any value of spin.

\section{A better way to discern particles}\label{SMSamend}

Muller and Saunders' Theorem 3 (539-40) contains the germ of a better way to secure discernment; i.e.~a way free of the criticisms discussed in Sections \ref{MullSau} and \ref{MullSee}.  This section develops the germ.  I proceed in stages.  First I outline the basic idea, and propose a relation which weakly and physically discerns two particles in any two-particle assembly, using statistical variance.  Then I investigate discernment for heterodox state spaces, in which particles may have definite position, and give a relation that will weakly in physically discern there too.  Finally, I propose a relation that weakly and physically discerns any two particles in an assembly of \emph{any} number of particles.

\subsection{The basic idea}\label{basicidea}

\noindent My  basic idea is that particles may be discerned by taking advantage of anti-correlations between single-particle states.  In the case of fermions, this is `easy' because of Pauli exclusion: in \emph{any} basis the occupation number for any single-particle state never exceeds one.  In the case of the other particles, it is more tricky, due to the fact that states for non-fermionic particles may have as terms product states with equal factors.   In these states, two or more particles are fully correlated, so there does not seem to be any quantum property \emph{or} relation which would discern them.  The solution is to change the basis to one in which anti-correlations appear with non-zero amplitude; the quantity associated with this new basis can then form the basis of a discerning relation.

Thus my strategy is discernment through anti-correlations, and the finding of anti-correlations through dispersion. For any state in which two particles are fully correlated, there will be dispersion in \emph{some other} basis; in particular, the dispersion will involve branches with non-zero-amplitudes in which the particles are anti-correlated.  

\subsection{The variance operator}\label{variance}

\noindent For simplicity, I focus exclusively on the two particle case.  We may take the assembly Hilbert space to be $L^2(\mathbb{R}^3)\otimes L^2(\mathbb{R}^3)$, but my results still carry over if we restrict to a symmetry sector, or add additional (e.g.~spin) degrees of freedom.  Anti-correlations between single-particle states in an eigenbasis for some single-particle quantity $A$ may be indicated by means of the following `standard deviation' operator:
\begin{equation}
\Delta_A := \frac{1}{2}\left(A\otimes\mathds{1} - \mathds{1}\otimes A\right) \ .
\end{equation}
Actually, I will use its square $\Delta_A^2$, the `variance' operator, which, like $\Delta_A$, is self-adjoint (since $A$ is).  Unlike $\Delta_A$, $\Delta_A^2$ \emph{is} a symmetric operator, so it is in line with the Indistinguishability Postulate (IP), and is therefore eligible to represent a physical quantity.  ($\Delta_A$ fails to be symmetric, since it is sent to \emph{minus} itself under a permutation.)

I also introduce the symmetric quantity $\overline{A}$, which may be viewed as the statistical mean of $A$, taken over the two particles:
\begin{equation}
\overline{A} := \frac{1}{2}\left(A\otimes\mathds{1} + \mathds{1}\otimes A\right) \ .
\end{equation}
Note that the over-line does \emph{not} indicate an expectation value: $\overline{A}$ is an operator.

By similarly defining $\overline{A^2} = \frac{1}{2}\left(A^2\otimes\mathds{1} + \mathds{1}\otimes A^2\right)$ we can express the variance operator more suggestively:
\begin{eqnarray}
\Delta_A^2 & = & \frac{1}{4}\left(A\otimes\mathds{1} - \mathds{1}\otimes A\right)^2\nonumber\\
&=& \frac{1}{4}\left(A^2\otimes\mathds{1} + \mathds{1}\otimes A^2 - 2A\otimes A\right)\nonumber\\
&=& \frac{1}{2}\left(\overline{A^2} - A\otimes A\right) \label{eq:DA1}
\end{eqnarray}
and
\begin{eqnarray}
\Delta_A^2 &=& \frac{1}{4}\left(A^2\otimes\mathds{1} + \mathds{1}\otimes A^2 - 2A\otimes A\right)\nonumber\\
&=& \frac{1}{2}\left(A^2\otimes\mathds{1} + \mathds{1}\otimes A^2\right) 
-  \frac{1}{4}\left(A^2\otimes\mathds{1} +2A\otimes A+ \mathds{1}\otimes A^2\right)\nonumber\\
&=& \overline{A^2} - \overline{A}^2\ . \label{eq:DA2}
\end{eqnarray}

It is the latter equation (\ref{eq:DA2}) which justifies the term `variance' for $\Delta_A^2$ and `standard deviation' for $\Delta_A$. But note again that it is not the ($c$-numbered) statistical variance of $A$ over a given wavefunction; it is the variance of the operator $A$ over the two particles: $\Delta_A^2$ is itself still an operator.  The former equation (\ref{eq:DA1}) makes it most clear that $\Delta_A^2$ measures the anti-correlation between each of the two particles' $A$-eigenstates.  In particular, for any state all of whose terms are product states with equal factors in the $A$-basis:
\begin{equation}
|\Psi\rangle = \sum_k c_k |\phi_k\rangle\otimes|\phi_k\rangle\ , \label{eq:cond}
\end{equation}
where
\begin{equation}
A|\phi_k\rangle = a_k|\phi_k\rangle\ ,
\end{equation}
it may be checked that the variance has eigenvalue zero.

In general, however, a state with anti-correlations will not be an eigenstate of $\Delta_A^2$.   For a generic state-vector
\begin{equation}
|\Phi\rangle = \sum_{ij} c_{ij} |\phi_i\rangle\otimes|\phi_j\rangle \label{eq:generic}
\end{equation}
we have
\begin{equation}
\Delta_A^2|\Phi\rangle = \frac{1}{4}\sum_{ij} c_{ij}\left(a_i - a_j\right)^2|\phi_i\rangle\otimes|\phi_j\rangle \label{eq:DAact}
\end{equation}
so that
\begin{eqnarray}
\Big\langle\Delta_A^2\Big\rangle &:=& \langle\Phi|\Delta_A^2|\Phi\rangle \nonumber\\
&=& \frac{1}{4}\sum_{ij} |c_{ij}|^2\left(a_i - a_j\right)^2\ . \label{eq:DAex}
\end{eqnarray}
If we assume that $A$ is non-degenerate (i.e.,~$a_i = a_j$ implies $i=j$), then it is clear from (\ref{eq:DAex}) that there is a positive contribution to the value of $\Big\langle\Delta_A^2\Big\rangle$ from \emph{every} anti-correlation that has a non-zero amplitude.

\subsection{Variance provides a discerning relation}

\noindent
If a two-particle state has anti-correlations in a single-particle quantity $A$, we can build a symmetric, irreflexive relation which discerns them.  The main idea is: if the expectation of the variance operator is non-zero, then this can be expressed as a relation between the two particles which neither particle bears to itself.

Following Muller \& Saunders (2008) and Muller \& Seevinck (2009), we define the operators
\begin{equation}
A^{(1)} := A \otimes \mathds{1}\ ; \qquad\qquad A^{(2)} := \mathds{1}\otimes A \ .
\end{equation}
These quantities, being non-symmetric, are unphysical, but they can be used to define physical quantities:
note, for example, that $\Delta_A \equiv \frac{1}{2}\left(A^{(1)} - A^{(2)}\right)$ and $\overline{A} \equiv \frac{1}{2}\left(A^{(1)} + A^{(2)}\right)$.  We then define the relation $R$ as follows:
\begin{equation}
R(A,x,y) \quad \mbox{iff} \quad \frac{1}{4}\left( A^{(x)} - A^{(y)}\right)^2\rho \neq 0 \ .
\end{equation}
In English: $R(A,x,y)$ holds for the state $\rho$ if and only if $\rho$ is not an eigenstate of the absolute difference between $x$'s and $y$'s operator $A$, with eigenvalue zero. Here the variable $A$ ranges over single-particle quantities and $x$ and $y$ range over particles.  This definition implies that
$R(A,1,2)$ {iff} $R(A,2,1)$, {iff} $\Delta^2_A\rho \neq 0$, and $\neg R(A,1,1) $ {and} $\neg R(A,2,2)$.
So $R(A,x,y)$ is symmetric and irreflexive for each $A$.  If $\Delta^2_A$ does not anihilate $\rho$, then we have $R(A,1,2)$ and $R(A,2,1)$; so in this case $R(A,x,y)$ weakly discerns particles 1 and 2.  Moreover, the discernment is categorical.

The question remains whether this discernment is physical.  I claim that it is, since the quantity $\frac{1}{4}\left( A^{(x)} - A^{(y)}\right)^2$, which is symmetric, can be understood as a measure of anti-correlations between $x$ and $y$ for the single-particle quantity $A$---i.e.,~a measure of difference between $x$'s and $y$'s values for $A$.  Thus it is no surprise that $\frac{1}{4}\left( A^{(x)} - A^{(y)}\right)^2 = 0$ for $x=y$; for no object can take a value for any quantity that is different from itself.  I claim that, so long as the single-particle operator $A$ has physical significance, so does $\frac{1}{4}\left( A^{(x)} - A^{(y)}\right)^2 = 0$.  I emphasize that the physical meaning of $\frac{1}{4}\left( A^{(x)} - A^{(y)}\right)^2 = 0$ should not be thought of as depending on $A^{(1)}$ or $A^{(2)}$'s having physical meaning.

There is an important analogy here with relative distance.  The relative distance between particle $x$ and particle $y$ need not be thought of as \emph{deriving its meaning from} the absolute positions of $x$ and $y$, even though the mathematical formalism of our theory may indeed allow us to define the relative distance in terms of these absolute positions.  We need not take these mathematical definitions as representative of any physical fact, since we are not forced to admit that an element of the theory's formalism which has a physical correlate also has physical correlates for all of its mathematical building blocks.  This is because these mathematical building blocks may contain redundant structure which is not transmitted to all of their by-products.  Such is the case of relative distance.  And in fact, relative distance is more than an analogy: for (squared) relative distance is an \emph{instance} of $\Delta^2_A$, if we set $A = \mathbf{Q}$, the single-particle position operator.

Note that an additional assumption is required to transmit physical significance from $\frac{1}{4}\left( A^{(x)} - A^{(y)}\right)^2 = 0$  to $R(A,x,y)$: we need to assume Muller and Seevinck's `strong property postulate'.  Recall that this states that any physical quantity of the assembly takes a certain value if and only if the assembly is in the appropriate eigenstate for that physical quantity's corresponding operator.  What is important here is the `only if' component of the biconditional: this enables us to say that the difference in  $x$'s and $y$'s values for $A$ is non-zero just in case the assembly is not in the eigenstate with eigenvalue zero---including when the assembly is not in an eigenstate at all.

I summarise the foregoing discussion in the following Lemmas:
\begin{description}
\item[Lemma 1] For all two-particle assemblies, and all single-particle quantities $A$, the relation $R(A,x,y)$ has physical significance if $A$ does, on the assumption of the strong property postulate.

\item[Lemma 2] For each state $\rho$ of an assembly of two particles, and each single-particle quantity $A$,  the relation $R(A,x,y)$ discerns particles 1 and 2 weakly, categorically and physically  if and only if $\Delta_A^2\rho \neq 0$, on the assumption of the strong property postulate.
\end{description}

\emph{Proofs:} See above. $\Box$

As with (SMS3), in the previous Section, we can instead forego the strong property postulate and instead take advantage of the Born rule, to settle for probabilistic discernment. To do so we define the relation $R'$ as follows:
\begin{equation}
R'(A,x,y) \quad \mbox{iff} \quad \frac{1}{4}\mbox{Tr}\left[\rho\left( A^{(x)} - A^{(y)}\right)^2\right] \neq 0 \ .
\end{equation}
Similar considerations to those above entail that $R'(A,1,2)$ {iff} $R'(A,2,1)$, {iff} $\left\langle\Delta^2_A\right\rangle \neq 0$.  And $\neg R(A,1,1) $ {and} $\neg R(A,2,2)$. So $R(A,x,y)$ weakly discerns particles 1 and 2 just in case $\left\langle\Delta^2_A\right\rangle \neq 0$.  Thus:
\begin{description}
\item[Lemma 3] For all single-particle quantities $A$, the relation $R'(A,x,y)$ has physical significance if $A$ does, on the assumption of the Born rule.

\item[Lemma 4] For each state $\rho$ of the assembly,  and each single-particle quantity $A$,  the relation $R'(A,x,y)$ discerns particles 1 and 2 weakly, probabilistically and physically,  if and only if $\left\langle\Delta_A^2\right\rangle \neq 0$ for that state.
\end{description}

\emph{Proofs:} See above. $\Box$

\subsection{Discernment for all two-particle states}

\noindent
So far we have seen that two particles in a state with non-zero variance in some single-particle quantity $A$---i.e.~two particles which are anti-correlated in $A$---may be discerned.  To guarantee discernment in all two-particle states it remains to be shown that, for \emph{any} such state, there \emph{will be} some single-particle quantity whose eigenbasis has anti-correlations.  In fact I prove a stronger result: namely that there is some single-particle quantity which discerns the two particles in all states of the assembly.  Moreover, this quantity is familiar: it is position; and since all particles have spatial degrees of freedom, it will be a quantity that will always be available to discern.

\begin{description}
\item[Theorem 1] For each state $\rho$ of an assembly of two particles,  the relation $R(\mathbf{Q},x,y)$ discerns particles 1 and 2 weakly, categorically and physically, where $\mathbf{Q}$ is  the single-particle position operator, on the assumption of the strong property postulate.
\end{description}

\emph{Proof:}
We assume the strong property postulate.  From Lemma 2, we know that $R(\mathbf{Q},x,y)$ discerns particles 1 and 2 weakly, categorically and physically, in the state $\rho$ if and only if $\Delta_\mathbf{Q}^2\rho \neq 0$.  Let us first consider only pure states, and later generalise to all states.

\emph{Pure states.}  Since we are working in the position representation, we use wavefunctions rather than state-vectors or density operators.  The most general form for the wavefunction of the assembly is
\begin{equation}
\Psi(\mathbf{x,y}) = \sum_{ij} c_{ij} \phi_i(\mathbf{x})\phi_j(\mathbf{y})\ ,
\end{equation}
where the $\phi_i$ are an orthonormal basis for $L^2(\mathbb{R}^3)$. (We assume zero spin, but the proof is trivially extended for any non-zero value for spin.) Now
\begin{eqnarray}
(\Delta_\mathbf{Q}^2\Psi)(\mathbf{x,y})&=& \sum_{ij} c_{ij}\left({x}^2\phi_i(\mathbf{x})\phi_j(\mathbf{y}) +\phi_i(\mathbf{x}){y}^2\phi_j(\mathbf{y}) - 2\mathbf{x}\phi_i(\mathbf{x}).\mathbf{y}\phi_j(\mathbf{y})\right) \nonumber\\
&=&  \left(\sum_{ij} c_{ij}\phi_i(\mathbf{x})\phi_j(\mathbf{y})\right)
\left(\mathbf{x-y} \right)^2 \nonumber\\
&=& \Psi(\mathbf{x,y}) (\mathbf{x-y})^2\ \label{Th1} 
\end{eqnarray}
(cf. Equation (\ref{eq:DAact})). This is the zero function only if $\Psi(\mathbf{x,y}) =0$ whenever $\mathbf{x \neq y}$. But then it cannot be represented in $L^2(\mathbb{R}^3) \otimes L^2(\mathbb{R}^3)$, since it is not a function.  Here I appeal to the fact that no wavefunction is ``infinitely peaked'' at the diagonal points of the configuration space.  (The necessary $\Psi$ can be written as a measure: $\Psi(\mathbf{x,y}) = f(\mathbf{x})\delta^{(3)}(\mathbf{x-y})$, for some function $f \in L^2(\mathbb{R}^3)$.  I return to this point in Theorem 3, below.) We may conclude that $(\Delta^2_\mathbf{Q}\Psi)(\mathbf{x,y}) \neq 0$.  It follows that $\Delta^2_\mathbf{Q}|\Psi\rangle\langle\Psi|\neq 0$.

\emph{Mixed states.}  We extend to density operators by taking convex combinations of (not necessarily othogonal) projectors.  We have that
\begin{equation} \label{Th1b}
\Delta^2_\mathbf{Q}\left(\sum_{i} p_i|\Psi_i\rangle\langle\Psi_i|\right) = 
\sum_{i} p_i\Delta^2_\mathbf{Q}|\Psi_i\rangle\langle\Psi_i| \neq 0
\end{equation}
since both the $p_i$ and the spectrum of $\Delta^2_\mathbf{Q}$ are positive. 

From Lemma 2, we conclude that $R(\mathbf{Q},x,y)$ discerns particles 1 and 2 weakly.  The discernment is categorical since we made no probabilistic assumptions.  Finally, the discernment is physical, as follows from  Lemma 1, the strong property postulate, and the fact that $\mathbf{Q}$ is physical.
$\Box$

We can now also prove
\begin{description}
\item[Theorem 2]  For each state $\rho$ of an assembly of two particles,  the relation $R'(\mathbf{Q},x,y)$ discerns particles 1 and 2 weakly, probabilistically and physically; where $\mathbf{Q}$ is the single-particle position operator; on the assumption of the Born rule.
\end{description}

\emph{Proof.}  We assume the Born rule. Then for any state $\rho$ we have (cf. Equations \ref{Th1}, \ref{Th1b}):
\begin{equation}
\left\langle\Delta^2_\mathbf{Q}\right\rangle = \sum_i p_i \int \textrm{d}^3\mathbf{x} \int\textrm{d}^3\mathbf{y}\ |\Psi_i(\mathbf{x,y})|^2 (\mathbf{x-y})^2\ ,
\end{equation}
which is always positive (cf.~Equation (\ref{eq:DAex})).  From Lemma 4, $R'(\mathbf{Q},x,y)$ therefore discerns weakly.  The discernment is probabilistic, since we assumed the Born rule.  Finally, the discernment is physical, as follows from  Lemma 3, the Born rule, and the fact that $\mathbf{Q}$ is physical. $\Box$

It may be objected against the proofs of the foregoing two Theorems that I rely too heavily on a technical feature of the assembly's Hilbert space, namely that it contains no states which exhibit no spread in $(\mathbf{x-y})^2$.  Effectively, unfavourable cases for discernment have been ruled out of the assembly's Hilbert space \emph{a priori}.  But this objection is easily dealt with. 

\begin{description}
\item[Theorem 3]  If we permit two-particle states to be represented by measures as well as by functions,  then for all such states, \emph{either}  $R(\mathbf{Q},x,y)$ \emph{or} $R(\mathbf{P},x,y)$ discerns particles 1 and 2 weakly, categorically and physically; where $\mathbf{Q}$ is the single-particle position operator, $\mathbf{P}$ is the single-particle momentum operator; on the assumption of the strong projection postulate.
\end{description}

\emph{Proof.}  The guiding idea is that any state will exhibit spread in either relative position or relative momentum, so no state is annihilated by both $\Delta^2_\mathbf{Q}$ and  $\Delta^2_\mathbf{P}$. 

We now allow measures, as well as functions, to represent states of the assembly.  Recall from the proof of Theorem 1 that $(\Delta_\mathbf{Q}^2\Psi)(\mathbf{x,y}) = 0$ only if $\Psi(\mathbf{x,y}) = 0$ whenever $\mathbf{x\neq y}$.  In this case  $\Psi(\mathbf{x,y}) = f(\mathbf{x})\delta^{(3)}(\mathbf{x-y})$, for some measure $f(\mathbf{x})$.  Note at this point that the two particles \emph{cannot be fermions}, since $\Psi(\mathbf{x,y}) = \Psi(\mathbf{y,x})$.
 We now move to the momentum basis by performing a Fourier transform on $\Psi$:
\begin{eqnarray}
\overline{\Psi}(\mathbf{k,l}) &=& \int \textrm{d}^3\mathbf{x} \int\textrm{d}^3\mathbf{y}\ \Psi(\mathbf{x,y}) e^{-i\mathbf{k.x}}e^{-i\mathbf{l.y}} \nonumber\\
&=& \int \textrm{d}^3\mathbf{x} \int\textrm{d}^3\mathbf{y}\ f(\mathbf{x})\delta^{(3)}(\mathbf{x-y})e^{-i(\mathbf{k.x +l.y})} \nonumber\\
&=& \int \textrm{d}^3\mathbf{x}\  f(\mathbf{x}) e^{-i\mathbf{(k+l).x}} \nonumber\\
&=& \overline{f}(\mathbf{k+l}).
\end{eqnarray}
This yields
\begin{equation}
(\Delta^2_\mathbf{P}\overline{\Psi})(\mathbf{k,l}) =  (\mathbf{k-l})^2\overline{f}(\mathbf{k+l})\ ,
\end{equation}
which is the zero function only if $\overline{f}(\mathbf{k+l}) = 0$ whenever $\mathbf{k \neq l}$.  But we can only satisfy this requirement if $\overline{f}$ is the zero function.  But in that case $\Psi(\mathbf{x,y})$ is the zero function, and so does not represent a state.  So if $(\Delta^2_\mathbf{Q}\Psi)(\mathbf{x,y})$ is the zero function, then $(\Delta^2_\mathbf{P}\overline{\Psi})(\mathbf{k,l})$ can't be. This result is easily extended to mixed states.

 With this result and Lemma 2 we conclude that either $R(\mathbf{Q},x,y)$ or $R(\mathbf{P},x,y)$  (or both) discerns particles 1 and 2 weakly.  The discernment is categorical, since we made no probabilistic assumptions.  Finally, the discernment is physical, given Lemma 1, the strong property postulate, and the fact that both $\mathbf{Q}$ and $\mathbf{P}$ are physical. $\Box$

It only remains to state
\begin{description}
\item[Theorem 4]  If we permit two-particle states to be represented by measures as well as by functions,  then for all such states, \emph{either}  $R(\mathbf{Q},x,y)$ \emph{or} $R(\mathbf{P},x,y)$ discerns particles 1 and 2 weakly, probabilistically and physically; where $\mathbf{Q}$ is the single-particle position operator, $\mathbf{P}$ is the single-particle momentum operator; on the assumption of the Born rule.
\end{description}

\emph{Proof:}  Left to the reader. $\Box$

So we have established the weak discernibility of indistinguishable particles in any two-particle assembly.  But my results are restricted to the two particle case.  Therefore, I  now turn to the many-particle case, and present theorems for assemblies of any number of particles.

\subsection{Discernment for all many-particle states}

\noindent
I begin by defining a generalized $N$-particle variance operator for each single-particle quantity $A$, for any $N\geqslant 2$.  For any single-particle quantity $A$, define
\begin{eqnarray}
\left(\Delta^{(N)}_A\right)^2 &:=& \overline{A^2} - \overline{A}^2 \nonumber\\
&=&\frac{1}{N}\sum_i^N \mathds{1}\otimes  \cdots  A^2_i \otimes \cdots \mathds{1}
- \left(\frac{1}{N}\sum_i^N \mathds{1}\otimes  \cdots  A_i \otimes \cdots \mathds{1}\right)^2
\nonumber\\
&=& \frac{1}{N^2}\sum^N_{i<j} \left(
\mathds{1}\otimes  \cdots  A_i \otimes \cdots \mathds{1}
\ - \
\mathds{1}\otimes  \cdots  A_j \otimes \cdots \mathds{1}\right)^2 \ . \label{eq:DAN}
\end{eqnarray}
Note that $\left(\Delta^{(2)}_A\right)^2 = \Delta^2_A$; cf.~Equations (\ref{eq:DA1}) and (\ref{eq:DA2}). 

Again, my strategy is to discern by setting $A=\mathbf{Q}$, the single-particle position operator.  If we act on any wavefunction $\Psi$ in $\bigotimes^N L^2(\mathbb{R}^3)$ with $\left(\Delta^{(N)}_\mathbf{Q}\right)^2$ we obtain, using (\ref{eq:DAN}),
\begin{equation} \label{DQN}
\left(\left(\Delta^{(N)}_\mathbf{Q}\right)^2\Psi\right)(\mathbf{x}_1, \ldots \mathbf{x}_N) = \frac{1}{N^2}\left(\sum_{i<j}^N(\mathbf{x}_i-\mathbf{x}_j)^2\right)\Psi(\mathbf{x}_1, \ldots \mathbf{x}_N) \ .
\end{equation} 
Now it is clear from Equation (\ref{DQN}) that we cannot proceed in the general case exactly as we did in the two-particle case.  That is: we cannot discern two particles---$a$ and $b$, say---by relying on the variance operator's annihilating the wavefunction.  For, the vanishing of the right-hand side of Equation (\ref{DQN}) is not a necessary condition for $a$ and $b$'s having vanishing relative distance: this relative distance may be zero, and yet there may still be non-zero contributions from the other particles.

However, we need only make mild adjustments to our previous strategy.  The idea is to look at regions of the configuration space for which $(\mathbf{x}_i - \mathbf{x}_j)^2$ is constant, except for when $i$ or $j$ equals $a$ or $b$.  We then indepedently vary $\mathbf{x}_a$ and $\mathbf{x}_b$.  If the wavefunction is non-zero for $\mathbf{x}_a \neq \mathbf{x}_b$, then we find variation in the right-hand side of Equation (\ref{DQN}) which can only be attributed to $a$ and $b$'s having non-vanishing relative distance---i.e.,~to their being discernible.

We first define a new dyadic relation between particles:
\noindent\begin{eqnarray} 
 &&\qquad\qquad\qquad D^{(N)}(x,y) \quad \mbox{iff} \nonumber\\
&&\!\!\!\!\!\!\!\!\!\!\!\!\!\!\!\!\!\!\!\!\!\!\!\!\!\!\!
\left(\left(\Delta^{(N)}_\mathbf{Q}\right)^2\Psi\right)(\mathbf{x}_1, \ldots \mathbf{x}_N)\ \neq
\ \frac{1}{N^2}
\!\!\!\!\!\!\!\!
\sum^N_{\scriptsize \begin{array}{c} i<j;\\ \langle i,j\rangle\neq\langle x,y\rangle;\\ \langle i,j\rangle\neq\langle y,x\rangle\end{array}}
\!\!\!\!\!\!\!\!
(\mathbf{x}_i-\mathbf{x}_j)^2\ \Psi(\mathbf{x}_1, \ldots \mathbf{x}_N)\ . \label{DDef}
\end{eqnarray} 
Note that $D^{(2)}(x,y)$  iff $R(\mathbf{Q},x,y)$; so $D^{(2)}$ is a physical relation.  Is $D^{(N)}$ a physical relation for \emph{any} $N$? First we note that the $N$-particle variance operator for position, $\left(\Delta^{(N)}_\mathbf{Q}\right)^2$, is a physical quantity, as is evident from its definition (\ref{eq:DAN}).  Now we need to make physical sense of the condition in the definition of $D^{(N)}$ (Equation (\ref{DDef})).

Recall that $R(A,x,y)$'s defining condition is to the effect that the wavefunction is not an eigenstate of the variance operator for some quantity (with eigenvalue zero); with the strong property postulate, this entails that the assembly does not have the corresponding physical property (namely, zero variance in that quantity).  Therefore, there can be no doubt that $R$'s defining condition is physically meaningful (so long as the strong property postulate is valid).  However, in the case of $D^{(N)}$, the condition is \emph{not} that $\Psi$ not be an eigenstate; the condition is rather that $\Psi$ not be sent to some specific function by the $N$-particle variance operator for position.  The strong property postulate is therefore no help in giving $D^{(N)}$'s defining condition physical significance.

We must settle for probabilistic discernment. $D^{(N)}$'s defining condition makes perfect physical sense if we assume the Born rule, since then the condition could be interpreted as the $N$-particle variance operator for position having an expectation value {not} equal to the value specified in the right-hand side of Equation (\ref{DDef}).  We can make this more explicit by defining another relation:
\begin{eqnarray} 
 &&\qquad\qquad\qquad\qquad D'^{(N)}(x,y) \quad \mbox{iff} \nonumber\\
&&\!\!\!\!\!\!\!\!\!\!\!\!\!\!\!\!\!\!\!\!\!\!\!\!\!\!\!
\left\langle\left(\Delta^{(N)}_\mathbf{Q}\right)^2\right\rangle\ \neq
\ \frac{1}{N^2}
\int \textrm{d}^3\mathbf{x}_1 \cdots \int \textrm{d}^3\mathbf{x}_N 
\!\!\!\!\!\!\!\!
\sum^N_{\scriptsize \begin{array}{c} i<j;\\ \langle i,j\rangle\neq\langle x,y\rangle;\\ \langle i,j\rangle\neq\langle y,x\rangle\end{array}}
\!\!\!\!\!\!\!\!
(\mathbf{x}_i-\mathbf{x}_j)^2\ |\Psi(\mathbf{x}_1, \ldots \mathbf{x}_N)|^2
\  . \label{D'Def}
\end{eqnarray} 

We may now prove
\begin{description}
\item [Thereom 5] For each state $\rho$ of an assembly of $N$ particles,  the relation $D'^{(N)}(x,y)$ discerns any two distinct particles $x$ and $y$ weakly, probabilistically and physically, on the assumption of the Born rule.
\end{description}

\emph{Proof.}  We prove this only for pure states and zero spin; the extension to mixed states and non-zero spin will be obvious, given our proof of Theorem 1.

It is clear that $\neg D'^{(N)}(x,x)$ for all $x$, since when $x=y$ the right-hand side of Equation (\ref{D'Def}) corresponds to the definition of the left-hand side (cf.~Equation (\ref{DQN})), and must therefore be equal to it.  Thus $D'^{(N)}$ is irreflexive.  To show that $D'^{(N)}$ discerns any two particles weakly, we need to prove that  $D'^{(N)}(x,y)$ holds whenever $x\neq y$.  

This we do by \emph{reductio}: assume that there are two particles $a$ and $b$ ($a \neq b$) for which $\neg D'^{(N)}(a, b)$.  Then we must have, by subtracting the right-hand side of Equation (\ref{D'Def}) from its left-hand side:
\begin{equation}
\frac{1}{N^2} \int \textrm{d}^3\mathbf{x}_1 \cdots \int \textrm{d}^3\mathbf{x}_N \
(\mathbf{x}_a-\mathbf{x}_b)^2\ |\Psi(\mathbf{x}_1, \ldots \mathbf{x}_N)|^2 = 0  \ .
\end{equation}
This holds only if $\Psi(\mathbf{x}_1, \ldots \mathbf{x}_N) = 0 $ whenever $\mathbf{x}_a \neq \mathbf{x}_b$.  So $$\Psi(\mathbf{x}_1, \ldots \mathbf{x}_N) = f(\mathbf{x}_1, \ldots \mathbf{x}_a, \ldots \mathbf{x}_{b-1}, \mathbf{x}_{b+1}, \ldots  \mathbf{x}_N)\delta^{(3)}(\mathbf{x}_a - \mathbf{x}_b),$$where $f$ is some $3(N-1)$-place function.  But then $\Psi$ is not a function, so it is not a state in $\bigotimes^N L^2(\mathbb{R}^3)$.  Thus $D'^{(N)}(a,b)$, and $D'^{(N)}$ weakly discerns any two particles in the assembly.

  The definition of $D'^{(N)}$ involves taking an expectation value, so it discerns probabilistically. Finally, the foregoing discussion establishes that $D'^{(N)}$ is a physical relation. $\Box$

Finally, I present 
\begin{description}
\item[Theorem 6]  If we permit states of an assembly of $N$ particles to be represented by measures as well as by functions,  then for all such states, \emph{either}  $D'(x,y)$ \emph{or} its momentum analogue discerns any two particles $x$ and $y$ weakly, probabilistically and physically, on the assumption of the Born rule.
\end{description}

\emph{Proof:} Left to the reader. The method is to carry over to the $N$-particle case the way in which proofs of Theorems 3 and 4 developed Theorems 1 and 2. $\Box$ 

\section{Conclusion}

Let me summarise the foregoing results.  A strong version of Leibniz's Principle of the Identity of Indiscernibles fails for all particles.  This version of the Principle permits discernment of two objects only by monadic properties, or relations to other objects not in the pair.  However, a weaker (and still non-trivial) version of the Principle is available, that was regimented by ($HB$) in Section 1.1.  According to this regimentation of the Principle, two particles may be discerned \emph{weakly}, i.e.~by some relation that applies \emph{between} the particles, but not reflexively to each.  This version of the Principle holds for \emph{all} particles: fermions, bosons and paraparticles.

 Previous attempts to establish this general result by Muller and Saunders (2008) and Muller and Seevinck (2009) have been seen to fail, due to their surreptitious use of mathematical predicates that can be given no physical interpretation.  However, physical predicates can be found which secure the result.  They derive their physical significance from the single-particle position operator (and, if needed, the single-particle momentum operator).  In the case of two-particle assemblies, this discernment may be categorical---that is, independent of all probabilistic assumptions---but we must assume the strong property postulate.  For assemblies of three of more particles, the discernment can only be probabilistic---that is, the Born rule must be assumed---but with that caveat the conclusions of Saunders, Muller and Seevinck are secured.

\section*{Acknowledgements}

Many thanks to Simon Saunders, Fred Muller and Michael Seevinck for several illuminating conversations on this topic.  I am also grateful to Steven French, James Ladyman, Mikl\'os R\'edei and audiences in London, Bristol and Cambridge.  A very large debt of gratitude is owed to Nick Huggett and Jeremy Butterfield for extensive comments, conversation and encouragement.  Finally, I am grateful to the Arts and Humanities Research Council and the Jacobsen Trust for their financial support during the writing of this paper.


\section*{References}

\quad\  Bassi, A. and Ghirardi, G. (2003), `Dynamical Reduction Models' {\em Physics Reports} {\bf 379}, p. 257-426.

Bell, J. S. (1964), `On the Einstein-Podolsky-Rosen paradox', \emph{Physics} \textbf{1}, pp.~195-200.

Bengtsson, I. and \.Zyczkowski, K. (2006), \emph{Geometry of Quantum States: An Introduction to Quantum Entanglement}.  Cambridge: Cambridge University Press.

Black, M. (1952), `The Identity of Indiscernibles', \emph{Mind} \textbf{61}, pp.~153-64. 

Butterfield, J.~N. (1993), `Interpretation and identity in quantum theory', \emph{Studies in the History and Philosophy of Science} \textbf{24}, pp. 443-76.

Caulton, A. and Butterfield, J. N. (2011), `Kinds of Discernibility in Logic and Metaphysics', forthcoming in the \emph{British Journal for the Philosophy of Science}.

Dieks, D. and Lubberdink, A. (2011), `How Classical Particles Emerge from the Quantum World', \emph{Foundations of Physics} \textbf{41}, pp.~1051-1064.

Earman, J. (ms.), `Understanding permutation invariance in quantum mechanics'. Unpublished manuscript.

Einstein, A., Podolsky B. and Rosen, N (1935), `Can Quantum-Mechanical Description of Physical Reality be Considered Complete?', \emph{Physical Review} \textbf{47}, pp.~777Ð780.

French, S. (2006), `Identity and Individuality in Quantum Theory', in E. N. Zalta (ed.), 
\emph{The Stanford Encyclopedia of Philosophy} (Spring 2007 Edition), $<$plato. stanford.edu/entries/qt-idind$>$. 

French, S. and Krause, D. (2006), \emph{Identity in Physics: A Historical, Philosophical and Formal Analysis.}  Oxford: Oxford University Press.

French, S. and Redhead, M. (1988), `Quantum physics and the identity of indiscernibles', \emph{British Journal for the Philosophy of Science} \textbf{39}, pp.~233-46.

Hilbert, D. and Bernays, P. (1934), \emph{Grundlagen der Mathematik}, Volume 1. Berlin: 
Springer. 

Huggett, N. (1999), `On the significance of permutation symmetry', \emph{British Journal for the Philosophy of Science} \textbf{50}, pp.~325-47.

Huggett, N. (2003), `Quarticles and the Identity of Indiscernibles', in K. Brading and E. Castellani (eds.), \emph{Symmetries in Physics: New Reflections}, Cambridge: Cambridge University Press, pp.~239-249.

Margenau, H. (1944), `The Exclusion Principle and its Philosophical Importance', 
\emph{Philosophy of Science} \textbf{11}, pp.~187-208. 

Massimi, M. (2001), `Exclusion Principle and the Identity of Indiscernibles: a Response to Margenau's Argument', \emph{The British Journal for the Philosophy of Science} \textbf{52}, pp.~303-330.

Messiah, A. and Greenberg, O. W. (1964), `The symmetrization postulate and its experimental foundation', \textit{Physical Review} \textbf{136}, pp.~248-67.

Muller, F.~A. and Saunders, S. (2008), `Discerning Fermions', \emph{British Journal for the Philosophy of Science}, \textbf{59}, pp.~499-548.

Muller, F. A. and Seevinck, M. (2009), `Discerning Elementary Particles', \emph{Philosophy of Science}, \textbf{76}, pp.~179-200.

Quine, W. V. O. (1960), \emph{Word and Object.} Cambridge: Harvard University Press.

Saunders, S. (2003a), `Physics and Leibniz's Principles', in K. Brading and E. Castellani 
(eds), \emph{Symmetries in Physics: Philosophical Reflections,} Cambridge: Cambridge 
University Press. 

Saunders, S. (2003b), `Indiscernibles, covariance and other symmetries: the case for non-reductive relationism', in A. Ashtkar, D. Howard, J. Renn, S. Sarkar and A. Shimony (eds.), \emph{Revisiting the Foundations of Relativistic Physics: Festschrift in Honour of John Stachel}, Amsterdam: Kluwer, pp.~289-307.

Saunders, S. (2006), `Are Quantum Particles Objects?', \emph{Analysis} \textbf{66}, pp.~52-63. 

Teller, P. (1986), `Relational Holism and Quantum Mechanics', \emph{The British Journal for the Philosophy of Science} \textbf{37}, pp.~71-81.

Weyl, H. (1928/1931), \emph{The Theory of Groups and Quantum Mechanics}. New York: Dover.

Weyl, H. (1949), \emph{Philosophy of Mathematics and Natural Science}.  Princeton: Princeton University Press.

\end{document}